# A Coarse-grained Deep Neural Network Model for Liquid Water

Tarak K Patra[1], Troy D. Loeffler[1], Henry Chan[1], Mathew J. Cherukara,[1] Badri Narayanan[2] and Subramanian K.R.S. Sankaranarayanan[1,3*]

[1]Center for Nanoscale Materials, Argonne National Laboratory, Argonne, IL 60439, USA
[2]Department of Mechanical Engineering, University of Louisville, Louisville, KY 40202, USA
[3]Department of Mechanical and Industrial Engineering, University of Illinois at Chicago, Chicago, IL 60607, USA

Corresponding author – skrssank@uic.edu, skrssank@anl.gov

**Abstract**

We introduce a coarse-grained deep neural network model (CG-DNN) for liquid water that utilizes 50 rotational and translational invariant coordinates, and is trained exclusively against energies of ~30,000 bulk water configurations. Our CG-DNN potential accurately predicts both the energies and molecular forces of water; within 0.9 meV/molecule and 54 meV/Å of a reference (coarse-grained bond-order potential) model. The CG-DNN water model also provides good prediction of several structural, thermodynamic, and temperature dependent properties of liquid water, with values close to that obtained from the reference model. More importantly, CG-DNN captures the well-known density anomaly of liquid water observed in experiments. Our work lays the groundwork for a scheme where existing empirical water models can be utilized to develop fully flexible neural network framework that can subsequently be trained against sparse data from high-fidelity albeit expensive beyond-DFT calculations.

**Key Words**: Water, Potential Energy Surface, Neural Networks, Thermodynamic properties, and Coarse-Grained Model

Molecular simulations are widely used in materials modeling for predicting structure-property relationships at the nanoscale. The accuracy of molecular simulations, such as molecular dynamics, hinges strongly on the quality of the inter-atomic potential used to describe the interactions between the atoms. Traditionally, classical molecular simulation relies on pre-defined functional forms to describe these atomistic interactions, which can often limit the chemistry and physics that can be captured. Recent work has shown that significant improvements can be made by using data driven approaches that employ extensive training data sets and advanced optimization.[1–3] However, it should be noted that there will always be a ceiling limit imposed by the use of pre-defined functional forms. Existing force-fields with pre-defined functional forms lack the flexibility and are not necessarily transferable from one material class to another (e.g. metals to oxides). There is clearly a need to incorporate flexibility in the functional form. In this respect, there is a lot of emphasis recently on neural network based potential models[4] – these models have become very popular owing to the rapid advancement in the computational resources/codes that allow for efficient generation of the training data (mostly first-principles based) and have simplified the subsequent training of the neural networks to describe the potential energy surface of materials.

Most of the existing deep neural networks (DNN) are based on all atom models. While the aim of these models is to retain first-principles accuracy at comparatively lower computational cost, they remain expensive compared to pre-defined functional forms. One of the ways to reduce the computational cost of these DNN models is coarse-graining. Traditionally, coarse-grained models tend to gain in efficiency by sacrificing accuracy. Recent efforts by Zhang and co-workers have focused on addressing this challenge.[5] The authors have introduced deep-learning coarse-grained potential (DeePCG) that constructs coarse-grained neural network trained with full atomistic data that preserves the natural symmetries of the system. Such models sample configurations of the coarse-grained variables accurately yet at much lower compute cost than the original atomistic model. They chose water as a representative system given the long-standing challenges it has posed to model developers world-wide.[1] In particular, liquid water exhibits maximum density at 277 K (which is 4 K above its melting point),[6] and modeling this anomalous behavior has been a long-standing problem. Many existing water models fail to capture such anomalies accurately.[7] While the DeePCG model is a promising tool and represents structure and dynamical properties of water reasonably, an accurate description of the thermodynamic anomalies (most notably density anomaly) has proven to be a challenge for coarse-grained DNN (CG-DNN) models.

One of the barriers to the development of CG-DNN models is the unavailability of large amount of high-quality data (from Quantum Monte Carlo or CCSD) that are essential for understanding a neural network topology, basis functions and parameterizations that are essential for capturing wide range of water properties for a wide range of environmental conditions. This challenge can be mitigated by transfer learning approaches where knowledge learned by different models (physics based and/or data driven model) can be built upon to develop a model that can capture wide range of properties and applications.[8] It utilizes knowledge gained during a training process for a second training thereby reducing the data required.[9,10]

Here, we aim to tackle the first step by developing a neural network model based on the knowledge of fixed functional form-based potential. In particular, we generate training data based on a recently developed coarse-grained bond order potential (CG-BOP) model[1] that capture temperature dependent properties of liquid water very accurately, including the density anomaly. Moreover, the BOP model is computationally very cheap and allows us to generate vast/diverse amount of training data. We train the network against only the energies of a large number of molecular configurations derived using the BOP water potential, and establish a machine learning framework where the deep neural network represent the potential energy surface of liquid water in its coarse-grained (CG) space using a set of rotational and translational symmetry functions.

Next, we interface the CG-DNN model with Monte Carlo (MC) simulations to evaluate the temperature-density correlation, while molecular dynamics (MD) simulations are used to compute transport properties (e.g., diffusivity). Even though the CG-DNN is parameterized only using the energies of configurations with varying number of water molecules, we find that the model accurately predicts forces as well as displays a maximum water density at ~280K which is close to its experimental value (277 K). Moreover, the predicted diffusivity of a water molecule at room temperature is also in reasonable agreement with experiments. As this neural network reasonably captures a wider range of liquid water properties, it can, in the future, serve as the basis for further training using a high-fidelity data set obtained from Quantum Monte Carlo (energies vs. configurations of water clusters). As techniques such as QMC are exorbitantly expensive and can only generate a sparse dataset, this CG-DNN model will allow the researchers to benefit from decades of works that went into fixed functional form-based water model development.

Table 1 Parameters of the 25 radial symmetry functions, $G^1$, that describe local physiochemical environment with in a cut-off distance $R_c = 3.5$Å. We chose the shift parameter $R_s$ to be 0.0.

| $G^1$ | $\eta$ (Å$^{-2}$) | $G^1$ | $\eta$ (Å$^{-2}$) | $G^1$ | $\eta$ (Å$^{-2}$) | $G^1$ | $\eta$ (Å$^{-2}$) | $G^1$ | $\eta$ (Å$^{-2}$) |
|---|---|---|---|---|---|---|---|---|---|
| 1 | 0.00417 | 6 | 0.01551 | 11 | 0.0576 | 16 | 0.21386 | 21 | 0.79406 |
| 2 | 0.00543 | 7 | 0.02016 | 12 | 0.07488 | 17 | 0.27802 | 22 | 1.03229 |
| 3 | 0.00706 | 8 | 0.02621 | 13 | 0.09734 | 18 | 0.36143 | 23 | 1.34197 |
| 4 | 0.00917 | 9 | 0.03408 | 14 | 0.12654 | 19 | 0.46986 | 24 | 1.74456 |
| 5 | 0.01193 | 10 | 0.04430 | 15 | 0.16451 | 20 | 0.61082 | 25 | 2.26790 |

Analogous to typical potential model development efforts, our procedure involves (1) defining or selecting a functional form (i.e., topology of DNN in this work), (2) constructing an extensive training data set using molecular simulations, (3) formulating a fitting procedure to optimize the weights or parameters of the network. Within this framework, our DNN consists of four layers of neurons; all the neurons/nodes of a layer are connected to every node in the next layer by weights in the manner of an acyclic graph. The two intermediate layers (hidden layers) consist of 10 nodes each. The input layer has 50 nodes which hold 50 symmetry functions that represent co-ordinates of the water's potential energy surface (PES). The output layer consists of one node that represents the potential energy of a water molecule in a given configuration. Besides, the input layer and the hidden layers contain a bias node that provides a constant signal to all the nodes of its next layer. The choice of this network topology is based on a large number of trials for capturing temperature dependent properties. Within this network topology, the three-dimensional Cartesian coordinate of the centroid of a water molecule $i$ in its liquid state is mapped into rotational and translational invariant co-ordinates as $G_i^1 = \sum_j e^{-\eta(r_{ij}-R_s)} \cdot f_c(r_{ij})$ and

$$G_i^2 = 2^{1-\zeta} \sum_{j,k \neq i}^{N_n} \left(1 + \lambda \cos\theta_{ijk}\right)^\zeta \cdot e^{-\eta(r_{ij}^2 + r_{ik}^2)} \cdot f_c(r_{ij}) \cdot f_c(r_{ik}),$$ where $f_c(r_{ij}) = 0.5 \cdot \left[\cos\left(\frac{\pi r_{ij}}{R_c}\right) + 1\right]$ for $r_{ij} < R_c$ and $f_c(r_{ij}) = 0.0$ otherwise. The indices $j$ and $k$ run over all the neighboring particles ($N_n$) within a cut-off distance $R_c = 3.5$Å. We have used 25 radial symmetry functions $G^1$ each with a distinct value of $\eta$, which are tabulated in Table 1. Similarly, 25 angular symmetry function $G^2$ are used, each with a distinct set of $(\eta, \zeta, \lambda)$ values. The parameters of these 25 angular

*symmetry functions are reported in Table 2.* The functional forms of these symmetry functions (Behler-Parrinello type symmetry functions) have been used successfully to construct PES of different molecular systems including water.[11–13]

To develop the CG-DNN model, we generate three sets of data: (1) training set containing energies of 30,000 CG liquid water configurations computed using BOP model (see Figure 1a for a distribution of these energies); (2) validation set containing energies of 3,000 configurations to provide an unbiased evaluation of a model fit on the training data, while tuning the DNN hyper-parameters; and (3) a test set of 10,000 configurations to assess the performance of the developed CG-DNN model. The number of molecules in each of these data sets varies from 2 to 128. Here, we employ the generalized representation of NNs for constructing the PES.[12] Within this representation, each molecule of a given configuration is represented by a NN. The total energy of a configuration is thus obtained as a sum of the molecular energies, defined as $E = \sum_{i=1}^{N_A} E_i$, where

Table 2: Parameters of the 25 angular Symmetry functions, $G^2$, that describe local chemical environment with in a cut-off distance $R_c = 3.5 Å$.

| $G^2$ | $\eta(Å^{-2})$ | $\lambda$ | $\zeta$ | $G^2$ | $\eta(Å^{-2})$ | $\lambda$ | $\zeta$ |
|---|---|---|---|---|---|---|---|
| 1 | 0.0004 | 1 | 2 | 14 | 0.0654 | 1 | 4 |
| 2 | 0.0054 | 1 | 2 | 15 | 0.0704 | 1 | 4 |
| 3 | 0.0104 | 1 | 2 | 16 | 0.0754 | -1 | 4 |
| 4 | 0.0154 | -1 | 2 | 17 | 0.0804 | -1 | 4 |
| 5 | 0.0204 | -1 | 2 | 18 | 0.0854 | -1 | 4 |
| 6 | 0.0254 | -1 | 2 | 19 | 0.0904 | 1 | 5 |
| 7 | 0.0304 | 1 | 3 | 20 | 0.0954 | 1 | 5 |
| 8 | 0.0354 | 1 | 3 | 21 | 0.1004 | 1 | 5 |
| 9 | 0.0404 | 1 | 3 | 22 | 0.1054 | -1 | 5 |
| 10 | 0.0454 | -1 | 3 | 23 | 0.1104 | -1 | 5 |
| 11 | 0.0504 | -1 | 3 | 24 | 0.1154 | -1 | 5 |
| 12 | 0.0554 | -1 | 3 | 25 | 0.1204 | 1 | 6 |
| 13 | 0.0604 | 1 | 4 | | | | |

$E_i$ is the output of the $i^{th}$ NN, and $N_A$ is the total number of molecules in a given configuration. We note that the architecture and weight parameters of all the molecular NNs are identical. During the training, the symmetry functions of each molecule of a configuration are fed to the corresponding NN *via* its input layer. In every NN, as shown schematically in Figure 1b, all the compute nodes in the hidden layers receive the weighted signals from all the nodes of its previous layer and feeds them forward to all the nodes of the next layer via an activation function as $x_{ij} = f\left(\sum_k W^i_{k,j} x_{i-1,k}\right)$. Here, $f(x) = \tanh(x)$ is used an activation function. As mentioned earlier, the sum of all the outputs from all the NNs serves as the predicted energy of the system. The error in the NNs, which is the difference between the predicted and reference energies of a given configuration, is propagated backward via the standard backpropagation algorithm.[14] All the weights that connect any two nodes are optimized using the Levenberg-Marquardt method[15] in order to minimize the error, as implemented within the framework of *aenet*[16] open-source code. All the configurations are sequentially fed to the NNs and the weights are optimized. The entire process is repeated until the error reaches a plateau. The mean absolute error (MAE) of the NNs is now defined as $MAE = \frac{1}{N}\sum_{\sigma=1}^{N}\left|E^{NN}(\sigma) - E^{\sigma}_{ref}\right|$, where $N$ is the total number of reference structures; $E^{NN}(\sigma)$ is the energy predicted by the NNs, and the actual energy of the reference configuration $\sigma$ is denoted as $E_{ref}^{\sigma}$. The MAE of the network during its training is shown as a function of training cycle in Figure 1c. We find that the error converges to a value of the order of $10^{-4}$ V within 50 cycles for both the training and validation data sets.

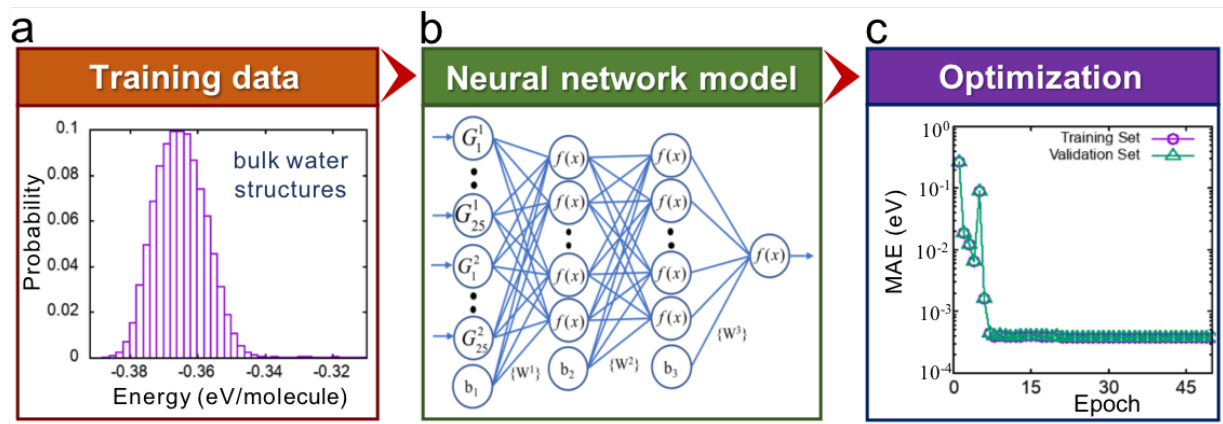

*Figure 1: Development of the CG-DNN model. (a) Data- the distribution of data used in training the neural network. (b) Model: Schematic representation of the neural network that takes 50 CG co-ordinates of a water molecule and output its energy. The input layer of the network holds these coordinates, which are represented by $G^1$ and $G^2$. The $G^1$ and $G^2$ are radial and angular symmetry functions, respectively. The network contains two hidden layers each of 10 compute nodes. All the compute nodes in the network are labeled as f(x). The $b_1$ and $b_2$ and $b_3$ are biased nodes which hold a constant value of one. The {$W^i$} represents the weights between nodes of $i^{th}$ and $(i+1)^{th}$ layers. (c) Optimization- Mean absolute energy during the training of the NN is plotted as a function of the training cycle for training and test data sets.*

The model is now used to predict the energy of the training and test sets. The predicted energy of the NN model is compared with the target values for the training and test sets in Figure 2a and 2b, respectively. The MAE in the training and test sets are 0.4 meV/molecule and 0.9 meV/molecule, respectively. This suggests that the neural network captures the CG-PES of liquid water quite accurately (within chemical accuracy), when compared to the reference BOP model. We next evaluate the predictive power the CG-DNN for properties that were not included in the training data set such as molecular forces. The force on a CG molecule is calculated by taking the derivative of its energy using the well-established chain rule.[11,17] The force on a CG particle $k$ along a Cartesian direction $r_{k,\alpha}$ where $\alpha = \{x,y,z\}$ can be written as

$$F_{k,\alpha} = -\frac{\partial E}{\partial r_{k,\alpha}} = -\sum_{i=1}^{N_A} \frac{\partial E_i}{\partial R_{k,\alpha}} = -\sum_{i=1}^{N_A} \sum_{j=1}^{M_i} \frac{\partial E_i}{\partial G_{i,j}} \frac{\partial G_{i,j}}{\partial R_{k,\alpha}}.$$

Here, the total energy of a molecular configuration $E$ is represented by the sum of environmental dependent atomic contribution $E_i$. The $N_A$ and $G_{i,j}$ are the number of atoms in a molecular configuration, and the $j^{th}$ symmetry function of particle $i$. The $M_i$ is the total number of symmetry function of particle $i$, which is 50 in this model. The correlation plots comparing the predicted force with the actual force based on the BOP model are shown in Figure 2c and 2d for the training and test data sets, respectively. We find the MAE in force prediction to be $\sim 53$ meV/Å. Although, the forces on the atoms are not explicitly included in the training data set, we find that the DNN predicted forces are in very good agreement with the reference forces (Fig. 2).

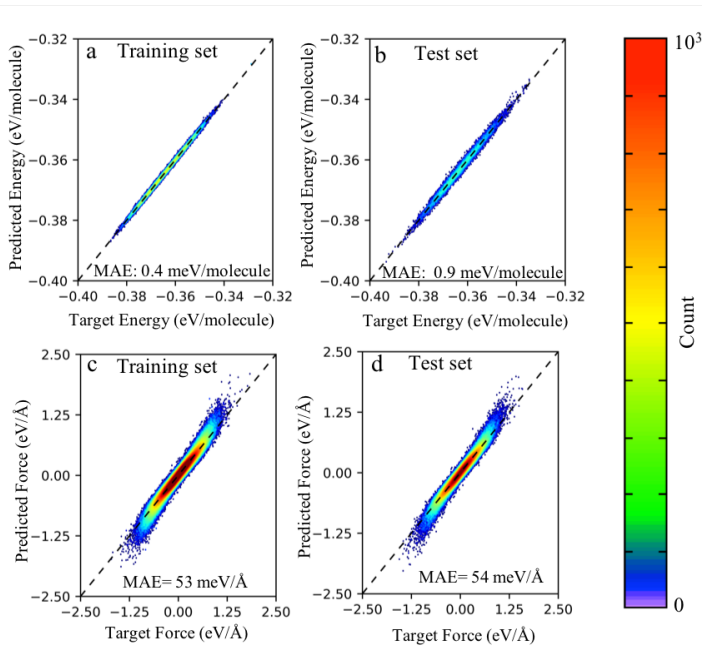

*Figure 2: Predictability of the CG-DN model. Energy correlations comparing DNN-prediction with the actual BOP energy for training (a) and test (b) sets that comprise 30000 and 10000 configurations, respectively. The dotted line represents the zero MAE. Force correlations are shown comparing actual force and that analytically derived from the DNN predicted energy for training (c) and test (d) sets. The color indicates number of molecular configurations at a specific value of energy (a,b) or force (c,d).*

Besides energies and forces, we have also validated the performance of the CG-DNN water model via temperature dependent properties of liquid water obtained from MC simulations performed under an isothermal isobaric (NPT) ensemble at P = 1 bar and T in the range of 250 K – 320 K. Each simulation consists of 1024 water molecules equilibrated for $10^5$ MC cycles in a periodic simulation box, followed by another $10^5$ MC cycles for the production run. At each temperature, values of the properties are averaged over sixteen independent production runs. Figure 3 shows a comparison of these properties between values predicted by CG-DNN and value predicted by BOP. Figure 3a shows the temperature dependent densities of liquid water predicted by CG-DNN and compared with BOP[1] and experimental[18] values. The CG-DNN predict density values within 0.003 gm/cc of BOP and experiment. For example, the CG-DNN predicts liquid water density of 0.998 gm/cc at T = 300 K whereas the BOP predicts a value of 0.9995 gm/cc. More importantly, the CG-DNN predicts the correct temperature - density correlation in liquid water (i.e., the density anomaly of liquid water). Moreover, the CG-DNN predicts maximum density at a temperature T ~ 280 K which is close to the expected values of T = 277 K. Next, we assess the structural predictions by evaluating the radial distribution function (RDF) and angular distribution function (ADF) of liquid water at T = 300 K. The Figure 3b and 3c show the comparison of RDF and ADF between the CG-DNN and BOP models, respectively. The CG-DNN predicts the RDF 1st, 2nd and 3rd peak positions (i.e., first, second and third coordination shells) to be at $r$ = 2.8 Å, 4.5 Å, and 6.8 Å, respectively, whereas ADF 1st and 2nd peak positions to be at $\theta$ ~ 47° and $\theta$ ~ 95°, respectively. These peak positions are all in excellent agreement with the BOP model as shown in Figure 3b and 3c. The RDF and ADF peak heights and widths of both models are also in excellent agreement. We note that both the DNN and BOP models predict RDF minimum to be at ~3.4 Å, which is deeper than that of the experimental RDF[19] as shown in Figure 3b. This might indicate over-structuring of liquid water, where the exchange of molecules between the first and second coordination shells is underestimated. However, the average coordination number of water molecules, *i.e.*, number of water neighbors in the first solvation shell, integrated out to the experimentally determined temperature independent isosbestic point (r = 3.25 Å), is 4.7 for BOP and DNN in accordance with the typical range of 4.3–4.7 observed in experiments.[19,20] Overall, the CG-DNN model reasonably captures the molecular structure of liquid water similar to that of the BOP model. We further conduct MD simulations in a NPT ensemble using the ASE open source package[21] to assess the dynamical properties predicted by the CG-DNN. Figure 3d shows block averages of the mean square displacement (MSD) of a water molecule at T = 300 K in the first 10 ps of a MD simulation. The water diffusivity calculated from the slope of the MSD curve is 3.60 x $10^{-5}$ $cm^2/s$, which is reasonably close to the value of 3.04 x $10^{-5}$ $cm^2/s$ predicted by the BOP model and the experimental value of 2.3 x $10^{-5}$ $cm^2/s$. [22]

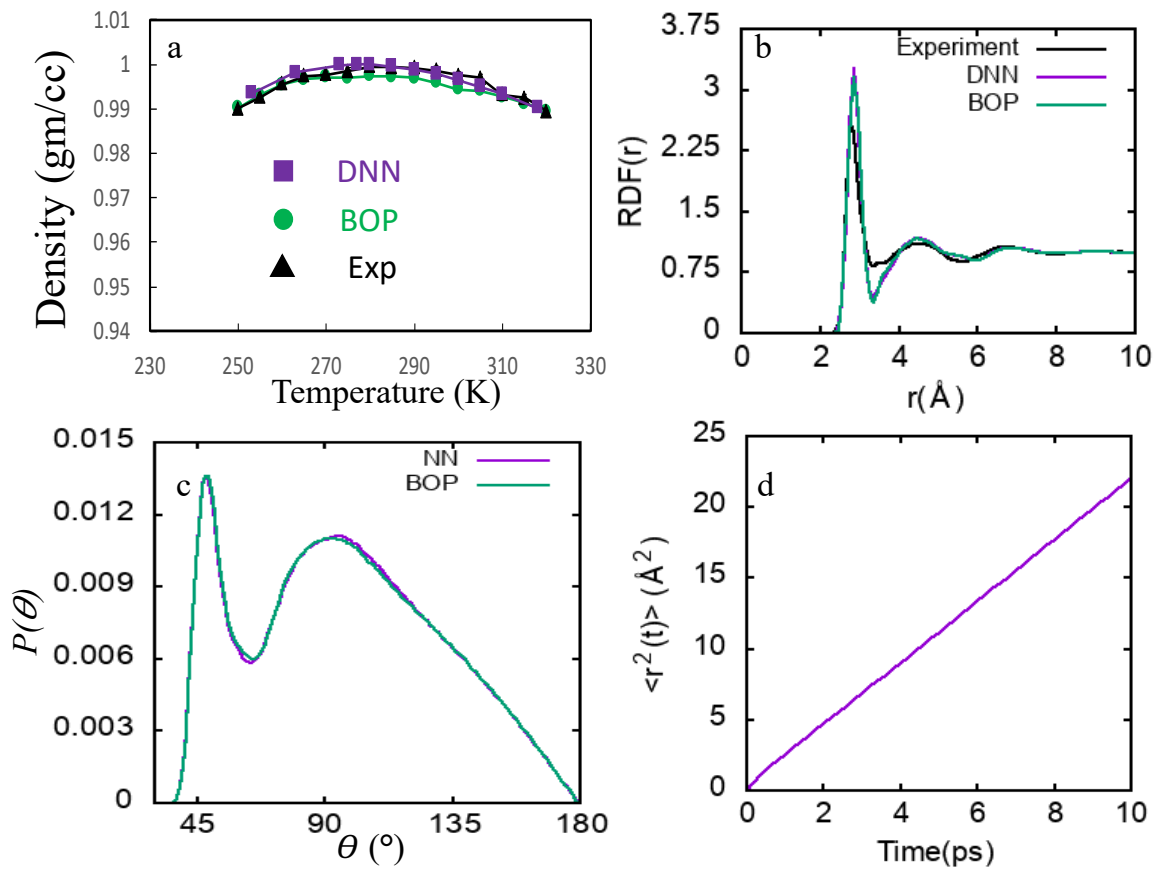

*Figure 3: Water properties computed via molecular simulation. The water density as a function of temperature is shown for DNN, BOP along with experimental data in (a). The radial distribution functions calculated based on DNN and BOP potentials are compared in (b) along with experimental data for temperature T=300 K. The (c) represents the angle distribution function P(θ) for both the DNN and BOP models for T=300K. Mean square displacement of a molecule (<r²(t)>) in liquid water for T=300 K  as computed via MD simulation of the DNN model is shown in (d).*

    In summary, the CG-DNN model accurately predicts the PES of coarse-grained liquid water and captures several important structural features and temperature-dependent properties (e.g. density anomaly) of liquid water. Unlike majority of the previous attempts where the development of DNN potentials were largely focused on constructing atomistic potential energy surface, here

we develop high dimensional representation of water in its coarse-grained space. Our approach is quite general and can be used to develop CG-DNN models for a wide variety of materials, especially soft-materials such as polymers. Traditionally, CG models have been used to describe the conformations of polymers *via* bead-spring type representations. Developing such phenomenological models comprises of optimizing the force-constants to match experimental observations or molecular data. The CG-DNN models can introduce flexibility into the functional form and overcome these limitations arising from being restricted to harmonic spring interactions by constructing a many-body representation for the bead interactions. In simulations of water, popular CG models such as mW and BOP allow for accelerated ice nucleation and growth studies since they do not explicitly treat hydrogen interactions, thereby eliminating constraints imposed by Pauling ice rules. Going forward, the current CG-DNN lays the groundwork for future efforts by water model developers to use this network as the starting point for transfer learning as well as generalizing this frame work to capture other water phases. we plan to extend the CG-DNN to describe the properties of ice polymorphs in addition to that of liquid water. We envision high-fidelity mesoscopic dynamical simulations to be facilitated by CG-DNN models.

## ACKNOWLEDGEMENTS

The authors would like to acknowledge the support from the Argonne National Laboratory through the Laboratory Directed Research and Development grant LDRD-2017-012-N0. Use of the Center for Nanoscale Materials, an Office of Science user facility, was supported by the U.S. Department of Energy, Office of Science, Office of Basic Energy Sciences, under Contract No. DE-AC02-06CH11357. This research used resources of the National Energy Research Scientific Computing Center, which is supported by the Office of Science of the U.S. Department of Energy under Contract No. DE-AC02-05CH11231. An award of computer time was provided by the Innovative and Novel Computational Impact on Theory and Experiment (INCITE) program of the Argonne Leadership Computing Facility at Argonne National Laboratory, which is supported by the Office of Science of the U.S. Department of Energy under contract DE-AC02-06CH11357.